\documentclass[doublecol]{epl2} 
\usepackage{amsmath}

\title{{Terahertz} out-of-plane 
{resonances} due to spin-orbit coupling}

\author{K. Morawetz\inst{1,2,3}}

\institute{                    
  \inst{1} M\"unster University of Applied Sciences,
Stegerwaldstrasse 39, 48565 Steinfurt, Germany\\
  \inst{2} International Institute of Physics (IIP),
Av. Odilon Gomes de Lima 1722, 59078-400 Natal, Brazil\\
\inst{3} Max-Planck-Institute for the Physics of Complex Systems, 01187 Dresden, Germany
}

\abstract{A microscopic kinetic theory is developed which allows to investigate non-Abelian SU(2) systems interacting with meanfields and spin-orbit coupling under magnetic fields in one, two, and three dimensions. The coupled kinetic equations for the scalar and spin components are presented and linearized with respect to an external electric field. The dynamical classical and quantum Hall effect are described in this way as well as the anomalous Hall effect where a new symmetric dynamical contribution to the conductivity is presented. The coupled density and spin response functions to an electric field are derived including arbitrary  magnetic fields. The magnetic field induces a staircase structure at frequencies of the Landau levels. It is found that for linear Dresselhaus and Rashba spin-orbit coupling a dynamical out-of-plane spin response appears at these Landau level frequencies establishing {terahertz resonances}.
}

\newcommand \be{\begin{eqnarray}}
\newcommand \ee{\end{eqnarray}}
\newcommand \ba{\begin{align}}
\newcommand \eea{\end{align}}
\newcommand {\ket}[1]{|#1\rangle}
\newcommand {\bra}[1]{\langle #1|}
\newcommand {\p}[1]{\partial_{#1}}
\newcommand \V{\vec}

\begin{document}

\maketitle

The spin-orbit coupling leads to the currently much debated spin-Hall effect \cite{SCH06} which was proposed \cite{DP71,DP71a} and first observed in bulk $n$-type semiconductors \cite{Ka04} and in 2D heavy-hole systems \cite{Wu05}. 
The extrinsic spin-Hall effect is due to spin-dependent scattering by mixing of spin and momentum eigenstates. The intrinsic effect is based on the momentum-dependent internal magnetic field created by the spin-orbit coupled band structure.{
}
Most observations are performed with the extrinsic \cite{Ka04,Sih:2005,VT06,SHMTI08} and only some for the intrinsic spin-hall effect \cite{Wu05,Br10}.  
A detailed discussion of possible occurring spin-orbit couplings in semiconductor bulk- and nanostructures can be found in \cite{WJW10} and in the book \cite{W03}. 
Recently even the spin-orbit coupled Bose-Einstein condensates have been realized \cite{LGS11}.

The spin-orbit coupling also induces a charge current perpendicular to the electric field due to the spin current as the anomalous Hall effect \cite{Si08,Nagaosa:2010} first described by \cite{Kar54} and measured by \cite{VT06}. The spin polarization takes over the role of a magnetic field. Finally the electrical current induced by the inhomogeneous spin density can be considered as inverse spin Hall effect \cite{av83,bak84}. 
{
}
{
}

Any spin-orbit coupling possess the general structure 
\be
\hat H^{\rm s.o.}=\V \tau\cdot \V b(\V k,\V R,T)=\tau \cdot (A,-B,C)
\label{so}
\ee 
like the Zeeman term 
with the Pauli matrices $\V {\bf \tau}$ momentum $\V k$, space $\V R$ and an also possible time-dependence $T$. There are different types of spin-orbit expansion schemes in the form (\ref{so}) in 2D and 3D \cite{Chen09} illustrated in table~\ref{ab}. 
The time reversal invariance of the spin current due to spin-orbit coupling 
requires that the coefficients $A(k)$ and $B(k)$ are odd functions of the momentum $k$ and therefore such couplings have no spatial inversion symmetry.

For direct gap cubic semiconductors such as GaAs the form (\ref{so}) of spin-orbit coupling arises by coupling of the s-type conductance band to p-type valence bands \cite{W03} and is six orders of magnitude larger than the Thomas term from the Dirac equation and has an opposite sign \cite{EHR05}. In a GaAs/AlGaAs quantum wells there can be two types of spin-orbit couplings that are linear in momentum. 
The linear Dresselhaus spin-orbit coupling is due to the bulk inversion asymmetry of the zinc-blende type lattice. It is proportional to the kinetic energy of the electron's out-of-plane motion and  decreases therefore quadratically with an increasing well width. 
The Rashba spin-orbit coupling arises from structure inversion asymmetry and it's strength can be tuned by a perpendicular electric field, e.g. by changing the doping imbalance on both sides of the quantum
 well. 
For the Rashba coupling and quadratic dispersion there has been shown that the 
spin-Hall effect vanishes \cite{Liu06,Kh06}.

\begin{table}
\begin{align*}
&
{\small
\begin{array}{ll@{\hspace{-0.5ex}}l@{\hspace{-0.5ex}}l@{\hspace{-0.5ex}}}
{\rm 2D-system} & A(k) & B(k) & C(k)\cr
\hline
{\rm Rashba} & \beta_R k_y & \beta_R k_x &\cr
{\rm Dresselhaus} [001] & \beta_D k_x & \beta_D k_y &\cr
{\rm Dresselhaus} [110] & \beta k_x & -\beta k_x &\cr
{\rm Rash.-Dressel.} & \beta_R k_y\!-\!\beta_D k_x & \beta_R k_x\!-\!\beta_D k_y &\cr
k^3\, {\rm Rashba (hole)} & i{\beta_R \over 2}(k_-^3\!-\!k_+^3)&{\beta_R \over 2}(k_-^3\!+\!k_+^3) &\cr
k^3\, {\rm  Dresselhaus} & \beta_D k_x k_y^2& \beta_D k_y k_x^2 &\cr
{\rm Wurtzite type} & (\alpha \!+\!\beta k^2 ) k_y & (\alpha\!+\!\beta k^2 ) k_x&\cr
{\rm graphene} & v k_x & -v k_y&\cr
{\rm bilayer graphene} & (k_-^2+k_+^2)/ 4m & (k_-^2-k_+^2)/ 4mi& 
\cr
{\rm 3D-system} &&& 
\cr
\hline
{\rm bulk}\, {\rm Dressel.} & k_x(k_y^2\!-\!k_z^2) & k_y(k_x^2\!-\!k_z^2) &k_z(k_x^2\!-\!k_y^2) \cr
{\rm Cooper pairs} & \Delta & 0 & {p^2\over 2 m} \!-\!\epsilon_F \cr
{\rm extrinsic} &&&\cr
 \beta={i\over \hbar}\lambda^2 V(k) &q_y k_z\!-\!q_z k_y&q_z
k_x\!-\!q_x k_z &q_x k_y\!-\!q_y k_x 
\end{array}
}
\end{align*}
\caption{\label{ab} Selected 2D and 3D systems with the Hamiltonian described by (\ref{sol}) taken from \cite{Chen09,cserti06} }
\end{table}

The present work is motivated by the observation of giant out-of-plane polarizations 
for in-plane fields first seen in InGaAs \cite{KMGA04}, With angle-resolved spectroscopy measurements a dominant out-of-plane spin component due to Rashba coupling was reported for Si-Au \cite{OMTMO10} and Bi thin films \cite{TSST11}. The numerical simulation reveals such an effect for sublattice asymmetry together with spin-orbit coupling  in graphene \cite{RKC11}. The optical out-of-plane spin polarization has been investigated in \cite{WP10} where a nonvanishing real and imaginary response was found in the THz regime and the spin dynamics has been treated within a reduced kinetic model in \cite{ERH07} restricted to high mobility. 

In order to understand this out-of-plane phenomenon in the whole frequency range including arbitrary magnetic fields we will derive the appropriate kinetic equations for impurity scattering on the meanfield level approximating the collisions by a relaxation-time and linearize them to get the dynamical density and spin response. Using this method, many-body lower-order approximations in the kinetic equations translate into higher-order response functions, e.g. the meanfield equation leads to GW (random phase approximation) and the Born approximation to the response with crossed diagrams etc. \cite{Mc02}. 

The formalism of nonequilibrium Green's function technique is used in the generalized Kadanoff-Baym notation introduced by Langreth and Wilkins \cite{LW72}. The two
independent real-time correlation functions for spin-$1/2$ fermions are
defined as 
\be 
G^>_{\alpha \beta}(1,2)=\langle \psi_\alpha
(1)\psi^\dagger_\beta (2) \rangle ,\,
G^<_{\alpha
\beta}(1,2)=\langle \psi^\dag_\beta (2)\psi_\alpha (1) \rangle .
\ee
Here, $\psi^\dagger$($\psi$) are the creation (annihilation) operators,
$\alpha$ and $ \beta$ are spin indices, and  numbers are cumulative
variables for space and time, $1\equiv(\V r_1,t_1)$. Accordingly, all the correlation functions without
explicit spin indices,  are understood as $2\times 2$ matrices in
spin space, and they can be written in the form $\hat C=C_0 +
\V \tau\cdot\V C $,  where $C_0$ ($\V C$) is the
scalar (vectorial) part of the function. This will result in
preservation of the quantum mechanical behavior concerning spin commutation
relations even after taking the quasi-classical limit  of the kinetic
equation. The latter one is obtained from the Kadanoff and Baym (KB) equation
\cite{KB62}  for the correlation function $G^<$
\ba \label{KB1}
-i(\hat G_R^{-1}\!\circ \hat G^<\!-\!\hat G^<\!\circ \hat G_R^{-1})=i(\hat G^R\!\circ \hat \Sigma^<\!-\!\hat \Sigma^<\! \circ \hat G^A) 
\end{align} 
where $\Sigma$ is the self-energy, and retarded
and advanced functions are defined as
$
C^{R/A}=\mp i\theta(\pm t_1\mp t_2)(C^>+C^<)+C^{HF}
$ 
with $C^{HF}$ the time-diagonal Hartree-Fock terms.
Products $\circ$ are
understood as integrations over intermediate  space and
time variables. We are interested in the $2\times 2$-matrix Wigner distribution 
function 
\be 
\hat \rho(\V p,\V R,T)&=&\hat G^<(\V p,\V R,t=0,T)
=f+\V \tau \cdot \V g
\label{FG}
\ee 
where we introduce mixed representation in terms of the center-of-mass variables $\V R=(\V r_1+\V r_2)/2$ and the Fourier transform of the relative variables $\V r=\V r_1-\V r_2\rightarrow \V p$ providing gauge invariance with the vector potential $A^\mu$ if the kinematical
momentum $k_\mu=p_\mu-e\int_{-1/2}^{1/2}{d\lambda A^\mu(X+\lambda
x)}$ is used. This treatment ensures that one has included all orders of a constant electric field \cite{bj91,Mo00}.
Compared to other formalisms with additional spin variable \cite{M88,ZMB10,LB10} we prefer here the separation of scalar and vector parts since they immediately give the distribution for density $n(q,T)=\sum_p  f$ 
and spin polarization $\V s(q,T)=\sum_p  \V g$. 

The spin-polarized Fermions interact with impurities of a spin-dependent potential $
V=V_0+\V \tau\cdot {\V V}
$ {where the vector part describes magnetic impurities or spin-dependent scattering responsible for zero-bias spin separation \cite{Gan06}}.  Due to the occurring product of the potential and the Wigner distribution (\ref{FG}), 
$
\hat V \hat \rho=V_0 \rho+{\V \rho} {\V V}+{\V \tau}\cdot [\rho {\V V}+V_0 {\V \rho}+i ({\V V}\times {\V \rho})],
$
the Hartree-meanfield selfenergy
\be
\hat \Sigma(p,R,T)=\sum\limits_{R'qQ}{\rm e}^{i q (R-R')}\hat \rho(Q+p,R',T) \hat V_{{q-Q\over 2},{q+Q\over 2}}
\label{Sig}
\ee
for intrinsic spin-orbit coupling possesses a scalar and a vector component
$
\Sigma_0(p,q,T)=
n(q) V_0(q)+\V s(q)\cdot \V V(q)$, 
$\V \Sigma_{\rm MF}(p,q,T)=
\V s(q) V_0(q)+n(q) \V V(q)$
.
This corresponds to an effective Hamiltonian with Fourier transformed $R\to q$
\be
H={k^2\over 2 m}+\Sigma_0(\V k,\V q,T)+e \Phi(\V q, T)+\V \tau \cdot \V \Sigma (\V k,\V q,T)
\ee
with
$\V {\Sigma}={\V \Sigma}_{MF}(\V k,\V q,T)+\V b(\V k,\V q,T)-\mu \V B$
the magnetic impurity meanfield, the spin-orbit coupling vector as well as the Zeeman term.
Extrinsic spin-orbit coupling requires 
convolutions with the particle current
$\V j(q,T)=\sum_p {\V p\over m} \, f
$
and the corresponding spin current
$\V d_{i}(q,T)=\sum_p {\V p \over m}\, [\V g]_i$ and creates a $\V k$-dependence.

We expand the $C=A\circ B$ in (\ref{KB1}) 
up to second-order gradients
and obtain the coupled equation system  for the scalar and vector components (\ref{FG}) by once forming the trace and once multiplying with $\V \tau$ and forming the trace 
\ba 
&{D_T} g_i\!+\!(\V {\p k}\Sigma_i\V {\p R}\!-\!\V {\p R}\Sigma_i\V {\p k}\!+\!(\V {\p k}\Sigma_i \times e\V B)\V {\p k} ) f\!
\!=\!{\!2 (\V \Sigma\!\times \!  \V g)_i}
\nonumber\\
&{D_T} f+(\V {\p k}\Sigma_i\V {\p R}-\V {\p R}\Sigma_i\V {\p k}+(\V {\p k}\Sigma_i \times e\V B)\V {\p k}) g_i \!=\!0
\label{kinet}
\end{align}
where $D_T=(\p T+\V {\cal F}\V {\p k}+\V v\V {\p R})$ describes the drift and force of the scalar and vector part with the effective velocity $\V {v}={\V k\over m}+\V {\p k} \Sigma_0$ and Lorentz force
${{\cal F}}=(e \V E - \V {\p R}
\Sigma_0+e \V {v} \times  \V B )$.

The second parts on the left side of (\ref{kinet}) represent the coupling between the spin parts of the Wigner distribution. The vector part contains additionally the spin-rotation term on the right hand side. 
These coupled mean field kinetic equations including the magnetic and electric field and spin-orbit coupling are one of the main results of the paper. 
{Without magnetic field, using linear spin-orbit coupling, and neglecting the meanfields one obtains the kinetic equation used so far in the literature, see \cite{TS07,D08} and references therein.} 

The stationary solution of the coupled kinetic equations (\ref{kinet}) has a two-band structure
$
\hat \rho({\hat \varepsilon})
= {f_++f_-\over 2}+\V \tau\cdot \V \xi \,\,{  f_+-f_-\over 2}
=f+\V \tau \cdot \V g
$
with the effective splitting $f_\pm=f_0(\epsilon_k(R)\pm|\V \Sigma(k,R)|)$, the selfconsistent meanfield $\epsilon_k(R)={k^2\over 2 m}+\Sigma_0(k,R)$, and precession $\V \xi(k,R)=\V \Sigma/|\V \Sigma|$.

Interestingly, for a homogeneous system without external magnetic field the kinetic equations (\ref{kinet}) decouple
\be
(\p t+e E\p k )f&=&0,\quad
(\p t+e E\p k )\V g=2(\V \Sigma \times \V g)
\label{homog}
\ee
and allow for a finite conductivity and Hall effect even without collisions. This is due to the
interference between the two bands and will be the reason for the following
anomalous Hall effect.  
Linearizing (\ref{homog}) and noting that $\V \Sigma\times \V g=0$ since $\V g=\V \xi (f_+-f_-)/2$ we obtain after Fourier transform of time the three terms [$E\partial_k=\V E\cdot \V \partial_k$]
\be
\!\!\delta \V g(\omega,k)\!=\! 
{i \omega  e E\partial_k\V g
\!-\!  {2\hbar} {\V  \Sigma} \!\times\! eE\partial_k \V g
\!-\!{4 i\over \omega}{\V  \Sigma} ({\V \Sigma} \cdot eE\partial_k \V g)
\over 4 |\Sigma|^2-\omega^2}.
\label{deltarho}
\ee
Each of these terms correspond to a specific precession motion analogously to the one in the conductivity of a charge in crossed electric and magnetic fields. To see this we note that with ${\V \Sigma} \leftrightarrow {\V B}$ and $\omega_c\leftrightarrow 2 |\V \Sigma|$ the time dependence of (\ref{deltarho}) is 
the solution of the Bloch-like equation 
$
m \dot {\V v}=e (\V v\times { \V B})+e \V E-m {\V v\over \tau_R}
$ for the current 
\ba
e n\V v(t)=\sigma_0 \int\limits_0^t \!{d\bar t\over \tau_R} &{\rm e}^{-{\bar t\over \tau_R}}\left \{ \! \cos{(\omega_c \bar t)}\V E_{t\!-\!\bar t}
\!+\! \sin{(\omega_c \bar t)}\V E_{t\!-\!\bar t}\times {\V B_0}
\right .\nonumber\\&\left .
\!+\![1\!-\!\cos{(\omega_c \bar t)}][\V E_{t\!-\!\bar t}\cdot {\V
  B_0}]{\V B_0}\right \},
\label{jtime}
\end{align}
{if we add a relaxation time approximation of the collisions $-\delta f/\tau_R$ to the right side of (\ref{kinet}) describing to dissipation $\omega^+= \omega+i/\tau_R$.} The solution (\ref{deltarho}) of the  
coupled equations (\ref{homog}) without collisions can also be obtained in two other ways, once in the helicity basis \cite{Liu05} and once from the Kubo Formula \cite{CB01,Sinitsyn:2007}. From (\ref{homog}) the particle and spin currents $\hat J_i=J_i+\V \tau\cdot \V S_i
=\sum_k [\hat \rho, v_i]_+
$
can be picked out after the corresponding trace, i.e. the charge current
\be
J_\alpha=2e\sum\limits_p \partial_\alpha \V \Sigma \delta \V g+2 e \sum\limits_p \p \alpha \epsilon \delta f=\sigma_{\alpha\beta}E_\beta
\ee
contains the spin-Hall effect $\sigma^{\rm as}$ as the third term of (\ref{deltarho}) and the first and second term will combine together to the symmetric part of the Hall conductivity 
\be
\left .
\begin{matrix}
\sigma_{\alpha\beta}^{\rm as}
\cr\cr
\sigma_{\alpha\beta}^{\rm sym}
\end{matrix} 
\right \}=e^2\sum\limits_k {g  \over
  1-{\omega^ 2\over 4|\Sigma|^2}}\,\left \{ \begin{matrix}
\V \xi\cdot (\partial_\alpha \V \xi\times \partial_\beta \V \xi)
\cr\cr
{i\omega\over
    2|\Sigma|}\partial_\alpha \V \xi\cdot  \partial_\beta \V \xi
\end{matrix} \right .
\label{cond}
\ee
with $\V \xi =\V \Sigma/|\Sigma|$. The symmetric $\sigma_{\alpha\beta}^{\rm sym}$ describes a part of the dynamical anomalous Hall conductivity which has been not presented in the literature yet. The part $\sigma_{\alpha\beta}^{\rm as}$ is the anomalous Hall conductivity as we can verify by the algebraic equivalence to the DC Hall conductivity from the Kubo formula \cite{Si08,Nagaosa:2010} for two bands $(\epsilon+\V \tau \cdot \V \Sigma)\ket \pm =\epsilon^\pm \ket \pm
$
\ba
\sigma_{\alpha\beta}^{\rm as}=-\epsilon_{\alpha\beta \gamma}{e^2}
\sum\limits_{n}\sum\limits_k f_n  (\V\partial_k \times { \V a_n})_\gamma
\end{align}
with Berry-phase connection \cite{Liu05,PhysRevB.84.075113}
$
{\V a_\pm}=i \hbar \bra \pm \V\partial_k\ket \pm=\hbar {\Sigma\pm\Sigma_z\over 2\Sigma} \V\partial_k \phi
$ for
$\Sigma_x-i \Sigma_y=\Sigma{\rm e}^{-i\phi}$.

In the second part we want to consider now the spin and density response to an external perturbing
electric field {$\delta \V E=-i \V q\Phi^{\rm ext}/e$} with arbitrary magnetic fields ${\V
  B}=B \V B_0=B \V e_z$. The linearized kinetic equations (\ref{kinet}) 
 using Bernstein's coordinates \cite{Be58} with $\V v(\phi)=(v \cos\phi,v\sin\phi,u)$,  $\Omega_t=\omega^+-\V q\cdot \V v(\phi)$, and $\phi ={\omega_c} t$, read
\begin{align}
&\left (-i\Omega_t-\p t\right ) \delta f+{i\V q\V {\p v}\Sigma_i\over m}\cdot \delta g_i=S_0\nonumber\\
&\left (-i\Omega_t-\p t\right ) \delta g_{i}+{i\V q\V {\p v}\Sigma_{i}\over m} \delta f-2 (\V \Sigma\times \V \delta g)_i=S_{i}
\label{lin}
\end{align}
where $\omega_c$ is the Lamor frequency, $\V \omega_c=\omega_c \V B_0$, and
\ba
&S_0=
{i\V q\V {\p v} f\over m} {\Phi^{\rm ext}}
\nonumber\\
&+
\left (i{\V q \over m}\!+\!\V \omega_c\times {\V {\p v^\Sigma} \over m}\right )
\delta \Sigma_0\V {\p v} f\!+\!
\left (i{\V q \over m}\!-\!\V \omega_c\times {\V {\p v^\Sigma} \over m}\right )
\delta \Sigma_{i}\V {\p v}g_{i}
\nonumber\\
&S_i=\left (i{\V q \over m}\!+\!\V \omega_c\!\times\! {\V {\p v^\Sigma} \over m}\right )\delta \Sigma_0\V {\p v} g_{i}\!+\!
\left (i{\V q \over m}\!-\!\V \omega_c\!\times\! {\V {\p v^\Sigma} \over m}\right )\delta \Sigma_{i}\V {\p v} f
\nonumber\\
&+{i\V q\V {\p v} g_{i}\over m} {\Phi^{\rm ext}}\!+\!2(\delta \V \Sigma\times \V g)_i
.
\label{source1}
\end{align}

Compared with the result without magnetic field we see that the source terms 
(\ref{source1}) get additional rotation terms coupled to the momentum-dependent 
variation of meanfields which is present only for extrinsic spin-orbit coupling. 
With $\delta \hat F=\delta f+\V \tau\cdot \delta \V g$ we rewrite (\ref{lin}) into 
\ba
-\p t \delta \hat F\!-\!i\Omega_t \delta \hat F\!+\!i(\V \tau \!\cdot\!  {D_+\V \Sigma})\delta \hat F\!+\!i\delta \hat F(\V \tau \!\cdot\! {D_-\V \Sigma})\!=\!\hat S
\label{linc}
\end{align}
where
{
$
D_\pm={\V q\cdot \V {\p v}/ 2 m}\pm 1$ and $\hat S=S_0+\V \tau\cdot \V S$.
Equation (\ref{linc}) is easily solved in quasiclassical, $D_\pm\V \Sigma \approx \pm \V\Sigma$, and relaxation time approximation, $\omega^+=\omega+i/\tau_R$
\be
\delta f\!+\!\V \tau\cdot \delta \V g=\int\limits_0^\infty d x
{\rm e}^{i (\omega^+ x-\V q R_x\V v)}{\rm e}^{-ix \V \tau\cdot \V \Sigma}\hat S_{t+x}{\rm e}^{i x \V \tau\cdot \V \Sigma}
\label{sol}
\ee
}
where the magnetic field enters the exponent as a matrix
\be
{\hat R_t}=\frac{1}{{\omega_c}}\left (\begin{matrix} 
{\sin{ \omega_c}t}&{1-\cos {\omega_c} t}&0\cr
{\cos{ \omega_c}t-1}&{\sin {\omega_c} t}&0\cr
0&0&{\omega_c} t
\end{matrix}
\right ).
\ee
Integrating (\ref{sol}) 
over momentum $\int {dk^\nu/ (2\pi\hbar)^\nu}$ for $\nu=3,2,1$ dimensions, respectively,
we obtain the density and spin-density response $\delta n\!+\!\V \tau\cdot \delta \V s$.
To work it out further one has
${\rm e}^{-i\V \tau \cdot \V \Sigma t}(S_0+\V \tau \cdot \V S){\rm e}^{+i \V \tau \cdot \V \Sigma t}=
S_0+(\V \tau\cdot \V S) \cos(2 t|\Sigma|)
-\V \tau (\V S\times \V \xi)\sin(2 t |\Sigma|)+{(\V \tau \cdot \V \xi)(\V S\cdot \V \xi)}(1-\cos(2 t |\Sigma|))
\approx S_0+\V \tau \cdot \V S-2\V \tau \cdot (\V S\times \V \Sigma)t
$
with the direction $\V \xi=\V \Sigma/|\Sigma|$.
The $\sin$ and $\cos$ terms are the different precessions analogously to (\ref{jtime}). For the limit of small $\Sigma$ it is sufficient to expand the $\cos$ and $\sin$ terms in first order.

Now we are in the position to see how the normal Hall and the quantum Hall effects are hidden in the theory. First we neglect any meanfield and spin-orbit coupling such that the $f$ and $\V g$ distributions decouple and consider the homogeneous $q\to 0$ limit to obtain from (\ref{sol})  and (\ref{source1})
\be
\delta f=  -{e\over m}\int\limits_0^{\infty} d t {\rm e}^{i (\omega+i/\tau_R) t
} \V {\p v} f(v_{\phi+\omega_c t})\cdot {\V E}
\label{Hallf}
\ee
where special care has been paid to the retardation since this provides the Hall effect which was overseen in many treatments of magnetized plasmas.
The charge current follows from (\ref{Hallf}) with $\sigma_0=n e^2 \tau_R/m$ by direct inspection
\ba
\V J &=e \int {d p^3\over (2 \pi \hbar)^3} \V {p(\phi)\over m}\, \delta f
%
=
\sigma_0 {1-i\omega \tau\over (1-i\omega \tau_R)^2+({\omega_c} \tau_R)^2 }
\nonumber\\
&
\times \!\left [\V E\!+\!{{\omega_c} \tau_R\over 1 \!-\!i \omega \tau_R} \V E\!\times\! \V B_0 
\!+\!{\omega_c^2 \tau_R^2\over (1 \!-\!i \omega \tau_R)^2}\V B_0 (\V B_0\cdot \V E)\right ]
\label{Jsol}
\end{align}
which appears again as the solution of the Bloch-like equation (\ref{jtime})
and the Hall conductivity
$\V J\sim \sigma_H \V B\times \V E $ follows 
\be
\sigma_H={R\bar \sigma^2\over 1+R^2 \bar \sigma^2 B^2},\, \bar \sigma={\sigma_0\over 1-i\omega \tau_R},\, R=-{1\over e n}. 
\ee

\begin{figure}
\includegraphics[width=4.2cm]{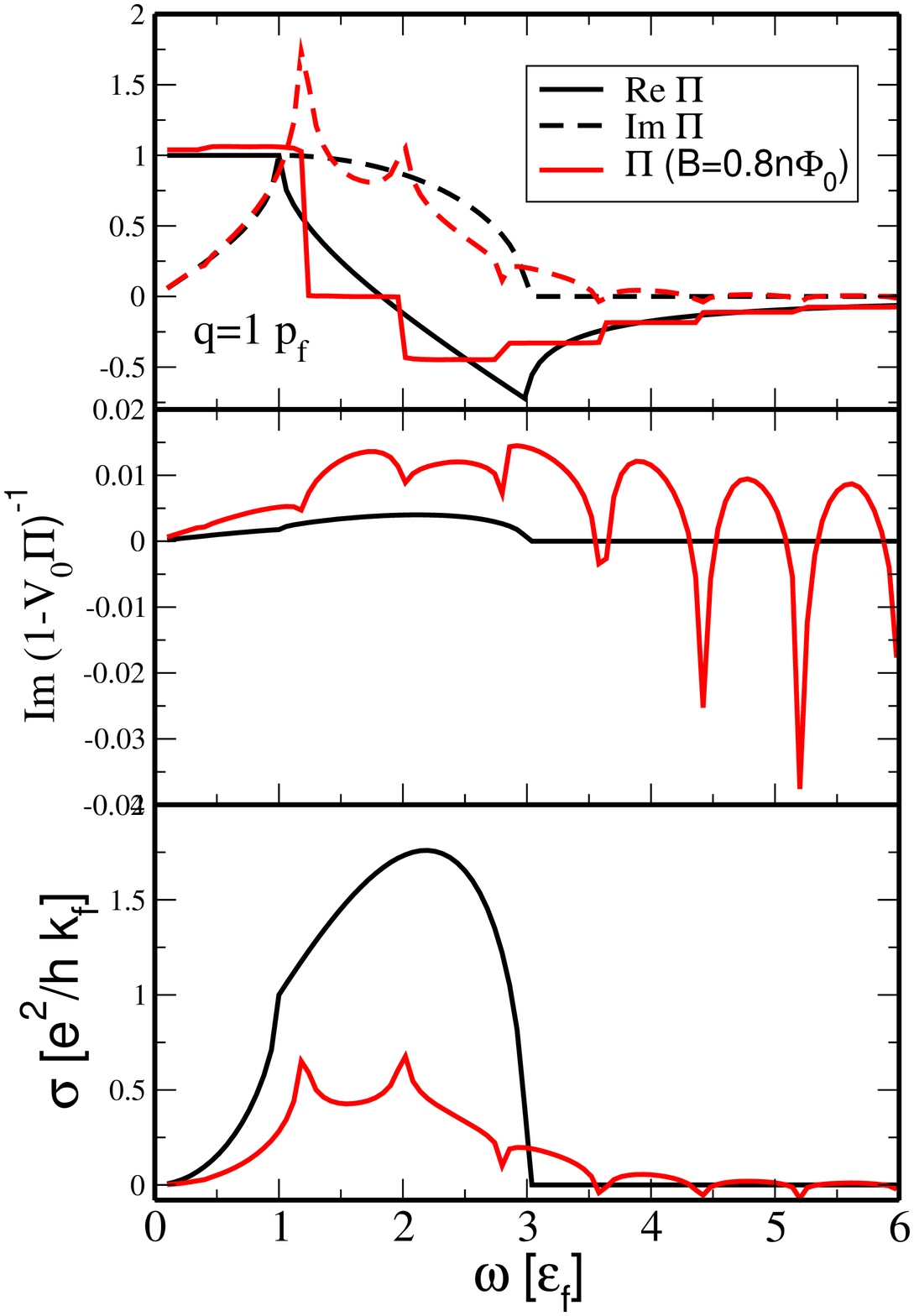}
\includegraphics[width=4.2cm]{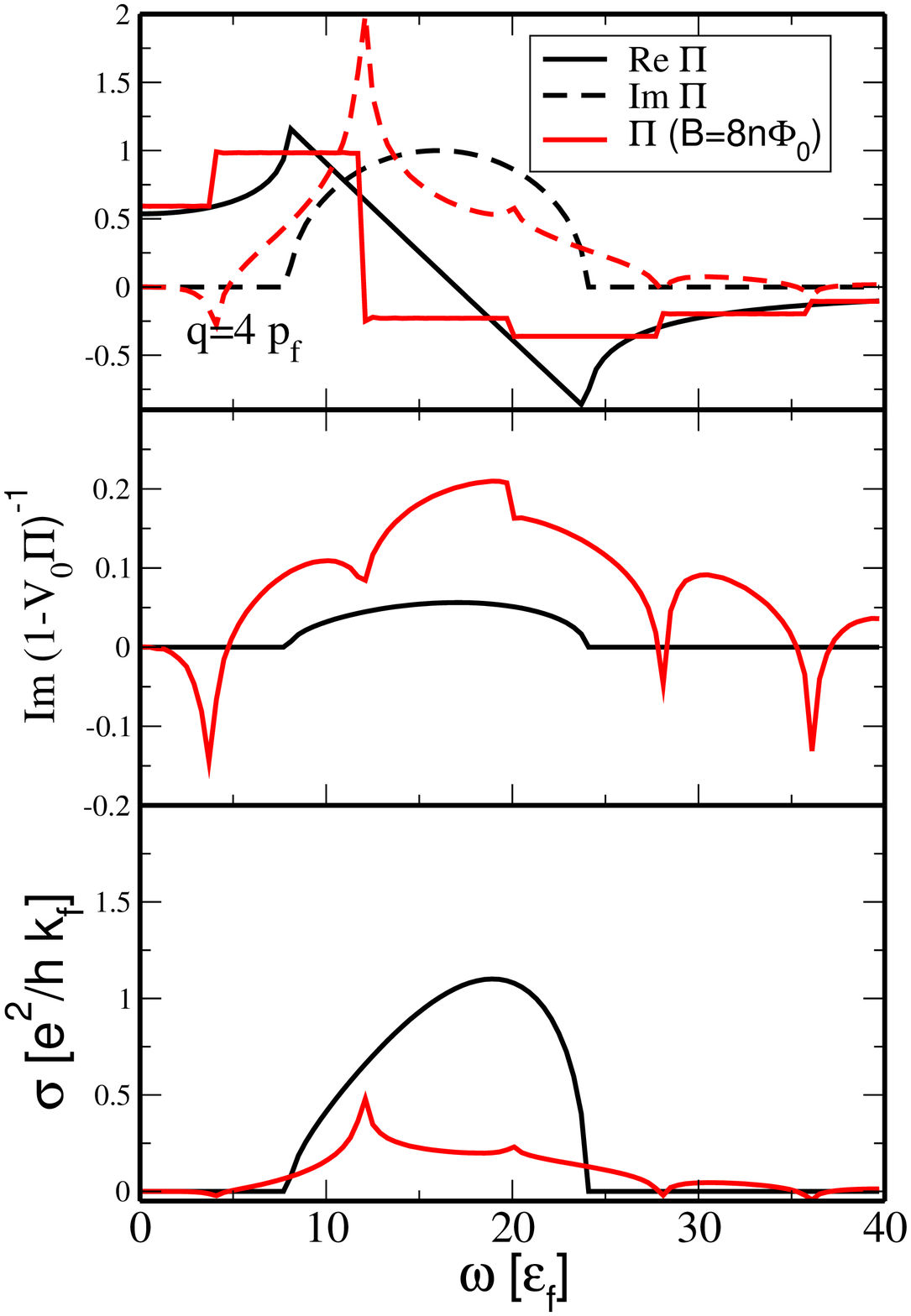}
\caption{The real (solid red) and imaginary part (dashed red) of the response function (above) together with the zero magnetic field ones (black lines), the excitation (middle) as well as dynamical conductivity (below) for two different wave vectors and magnetic fields ($\Phi_0=h/2e$) for a quasi-2D electron system with charged-impurity density $7\times 10^{10}$cm$^{-2}$. \label{magn}}
\end{figure}

Next we consider low temperatures such that the motion of electrons becomes quantized in Landau levels and we have to use the quantum kinetic equation and not the quasiclassical one. Linearizing the quantum-Vlassov equation, $\dot \rho-\frac i \hbar [\rho, H]=0$, one obtains
\be
\delta \rho_{nn'}=-e\V E \cdot \V x_{nn'} {\rho_n-\rho_{n'}\over \hbar \omega -E_n+E_{n'}}
\ee  
in the discrete basis $\bra n\leftrightarrow \bra {p+\frac q 2},\quad \ket {n'} \leftrightarrow \ket {-p+\frac q 2}$. Compared with the quasi-classical equation we see that our quasiclassical results can be translated into the quantum result by applying the rule
\be
\V E\cdot \V {\p p} f\to \V E\cdot \V v_{nn'}{f_n-f_{n'}\over E_{n'}-E_n}.
\ee
For the static, $\omega=0$, Hall conductivity (\ref{cond}) we get, e.g.
\be
\sigma_{\alpha\beta}={e^2\hbar i\over L_yL_z}\sum\limits_{nn'} f_n (1-f_{n'}) {1-{\rm e}^{\beta (E_n-E_{n'})}\over (E_n-E_{n'})^2} v_{nn'}^\alpha v_{n'n}^\beta
\label{si1}
\ee
which is exactly the result of the Kubo formula. The further evaluation has been performed by Vasilopoulos \cite{V85,VV88} and 
{one arrives for $T\to 0$ at the quantum Hall result $\sigma_{\alpha\beta}\to {e^2\over h}(n+1)$.}

Now that we have gained trust in the coupled equation (\ref{sol}) we calculate the general response.
Let us consider only the meanfield due to impurity scattering and no extrinsic spin-orbit coupling. 
{The resulting linear system for the particle and spin density response from (\ref{sol}) reads} 
\ba
&(1-\Pi_0 V_0-\V \Pi \cdot \V V){\delta n}=\Pi_0 {\Phi^{\rm ext}}
+(\Pi_0\V V+\V \Pi V_0) \cdot {\delta \V s}
\nonumber\\
&(1\!-\!\Pi_0 V_0){\delta \V s}=\V \Pi_3 {\Phi^{\rm ext}}
\!+\!(V_0 \V \Pi_3\!+\!\Pi_0 \V V\!+\!\V V\times \V \Pi_2\!+\!\V \Pi_{\V \xi}\cdot \V V){\delta n}
\nonumber\\
&
\qquad \qquad +\V \Pi_3(\V V \cdot {\delta \V s})+V_0\V \Pi_{\V \xi}\cdot {\delta \V s}
+V_0\V \Pi_2\times {\delta \V s}
\label{sys}
\end{align}
with the abbreviations $\V \Pi_2=\V \Pi_g-\V \Pi_{xf}$
and
$\V \Pi_3=\V \Pi+\V \Pi_{xg}$. One recognizes how the spin excitation influences the density response and vice versa. The different magnetfield-dependent polarizations with $\V g=\V \xi g$ read
\be
\left .\begin{matrix}
\Pi_0 \cr
\V \Pi\cr
\V \Pi_{g}\cr
\V \Pi_{xf}\cr
\V \Pi_{xg}\cr
\V \Pi_{\V \xi}\cr
\end{matrix}
\right \}
=\sum\limits_p \int\limits_0^{\infty} d t {\rm e}^{i (\omega^+ t-\V q \cdot \hat  R_{t}  \cdot \V v)}
\left \{
\begin{matrix}
i \Delta f \cr
iq \Delta \V g\cr
2 \V g\cr
2 i t {\V \Sigma} \Delta f\cr
2 i t {\V \Sigma} \times \Delta \V g\cr
4 t \Sigma g (1-\V \xi\circ \V \xi)
\end{matrix}
\right ..
\label{form}
\ee
with $\Delta f=q \p p f$ for quasiclassical kinetic equation and  $\Delta f=f_{p+\frac q 2}-f_{p-\frac q 2}$ for quantum-Vlassow respectively.
Other approaches to the response consider special expansions in magnetic field or interaction strength \cite{Shiz07,KB09,ZMB10,LB10}. We suggest that (\ref{sys}) is the general result for the coupled spin-density response and the main result of the paper. 
It can be best understood by different special cases. Neglecting the spin components one obtains the magnetic field-dependent density response $\delta n=\kappa {\Phi^{\rm ext}}$ corresponding to GW (RPA, rainbow, ...) approximation, the dynamical conductivity, and dielectric function
\be
\!\!\kappa(q\omega)\!=\!{\Pi_0(q,\omega)\over 1\!-\!V_0(q)\Pi_0(q,\omega)},\, \sigma\!=\!-i \omega \epsilon_0 \epsilon,\, \epsilon\!=\!1\!-\!V_0 \Pi_0
\ee
which is exactly the quasiclassical response function in magnetic
fields \cite{WZT05} first derived by Bernstein \cite{Be58}. It reduces to 
the Lindhard response for vanishing magnetic fields $\omega_c\to 0$.

The effect of the magnetic field we see in figure \ref{magn}, where we present as an exploratory example the real and imaginary parts of polarization for a  quasi-2D electron system \cite{Mc02}, the collective excitation spectrum ${\rm Im} (1-V_0 \Pi_0)^{-1}$, and the conductivity for zero temperature. One nicely recognizes that at the frequencies of the Landau levels $\omega=\omega_c (n+\frac 1 2)$ the conductivity shows peaks according to the staircase behavior of the real part of polarization since we used the quantum kinetic result $q\cdot \p p f\to f(p+\frac 1 2 q)-f(p-\frac 1 2 q)$.

\begin{figure}
\includegraphics[width=8cm]{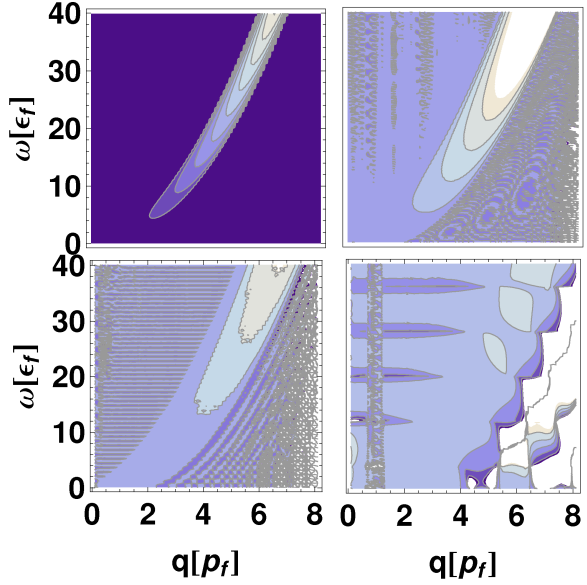}
\caption{The excitation spectrum of left Fig. \ref{magn} for $B=0,0.08\Phi_0$n (above) and  $B=0.8,8\Phi_0$n (below).\label{exo0_01}}
\end{figure}

The collective excitation spectrum in figures \ref{exo0_01} illustrates the position and width of the collective density mode. One sees how with larger magnetic field the Landau levels create excitation modes independent of the wave vector on the left side. The classical Bernstein modes are visible for higher magnetic fields as parallel acoustic modes right to the main line. Interestingly a gap in the main mode appears for very high fields, where excitations are suppressed by the magnetic field.

As a next special case we neglect the meanfield only keeping the spin-orbit coupling and the magnetic field $V_0=\V V=0$ to obtain a decoupled response
$
\delta n=\Pi_0 {\Phi^{\rm ext}},
\delta \V s=(\V \Pi+\V \Pi_{x g}){\Phi^{\rm ext}}
$. This shows that the density response is unaffected by the spin excitation and only modified due to the magnetic field when no meanfields are present. The spin response are explicitly dependent on the spin-orbit coupling due to $\V \Pi_{xg}$ of (\ref{form}) which is the response corresponding to the anomalous Hall effect.

\begin{figure}
\includegraphics[width=8.5cm]{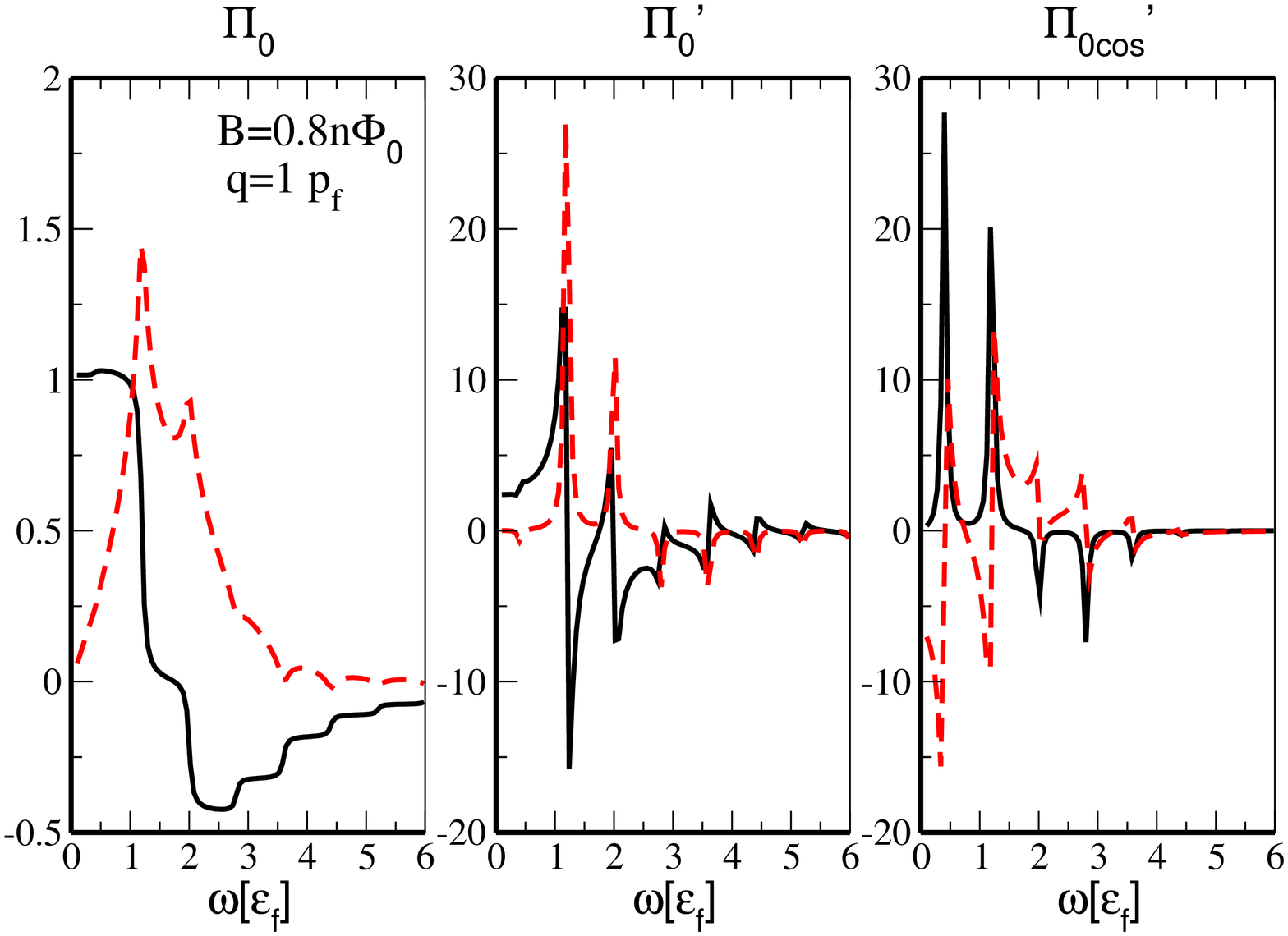}
\caption{The three occurring polarization functions of (\ref{pola}) according to left Fig. \ref{magn} with real parts (black solid line) and imaginary parts (red dashed line).\label{pol}}
\end{figure}

Now we return to the general case (\ref{sys}) 
where we neglect the spin-flip mechanism $\V V=0$ (magnetic impurities) such that the equation system reduces remarkably. Restricting here to first order in $V_0$ one gets
 \be
\delta \V s&=&\left ({\V \Pi_3\over 1-2 V_0\Pi_0}-V_0 \V \Pi_2\times \V \Pi_3+ V_0 \V \Pi_{\V \xi} \cdot \V \Pi_3\right ){\Phi^{\rm ext}}\nonumber\\
\delta n&=&\left ( {\Pi_0 \over 1-2 V_0\Pi_0}+V_0 (\V \Pi \cdot \V \Pi_3-\Pi_0^2)\right ){\Phi^{\rm ext}}.
\ee
The spin excitation influences the density response only if the vector distribution has a different momentum distribution in different directions (anomalous Hall effect). Lets consider the simpler case where the spin distribution has the same momentum distribution in each direction, $\V g=\V \xi g$. In 2D with $\V s=(s_x^0,s_y^0,0)=\V \xi |\V s|$ follows $\V \Pi_3 \cdot \V \Pi=\V \Pi^2=\Pi_0^2$
and
 \be
\delta \V s&=&{\V \Pi_3\over 1-2 V_o\Pi_o}{\Phi^{\rm ext}},\qquad
\delta n={\Pi_0\over 1-2 V_o\Pi_o}{\Phi}^{\rm ext}.
\ee
The corresponding spin polarization for Dresselhaus (Rashba) linear spin-orbit coupling
\be
\V \Pi_3\!=\!\left (\begin{matrix}s_x^0\cr s_y^0\cr 0\end{matrix}\right )\Pi_0\!+\!
\mu \! \left (\begin{matrix}s_x^0\cr s_y^0\cr 0\end{matrix}\right )\!\!\times\!\! \V B\,
\Pi_0'\!-\!\!\left (\begin{matrix}0\cr 0\cr \beta_D s_y^0 \!-\!\beta_R 
    s_x^0\end{matrix}\right ) {\Pi_{0cos}'}
\label{pola}
\ee
shows an out-of-plane polarization due to the induced spin precession.
The three different occurring polarization functions are plotted in Fig. \ref{pol} where $\Pi_0'$ indicates the frequency derivative.

To summarize, we have derived the density and spin response to an external electric field including arbitrary magnetic fields and spin-orbit coupling from the coupled kinetic equations for scalar and spinor distributions. The spin and density components are coupled due to meanfield interactions. Besides the dynamical classical Hall and quantum Hall effects also the anomalous Hall effect follows from the kinetic equations. A new frequency-dependent term in the anomalous Hall conductivity is presented. The spin-orbit coupling leads to three different precession motions. The magnetic field induces a staircase structure in the frequency dependence. For a linear Dresselhaus and Rashba coupling a spin response component out-of plane appears as frequency derivatives. Therefore it becomes large around these 
frequency steps and provides a sign change in the optical conductance as observed \cite{TSST11} and explains why this effect is present in the clean systems only in the dynamical case \cite{WP10}. 
Analyzing these frequencies it turns out that they appear on the values of the Landau levels which are in the {terahertz} regime for typical GaAs semiconductors \cite{WP10}. In other words the spin-orbit coupling suggests that under high magnetic fields, out-of-plane spin {terahertz resonances should be generated}. 

\end{document}